\documentclass[graybox]{svmult}

\usepackage{type1cm}        
\usepackage{makeidx}         
\usepackage{graphicx}        
\usepackage{multicol}        
\usepackage[bottom]{footmisc}

\usepackage{amsmath,amssymb}

\newcommand{\EE}{E}

\makeindex             

\begin{document}

\title*{Revisiting the theoretical basis of agent-based models for pedestrian dynamics}
\titlerunning{Revisiting agent-based models for pedestrians}
\author{Iñaki ECHEVERRÍA-HUARTE \and Alexandre NICOLAS}
\authorrunning{Echeverr\'ia-Huarte \and Nicolas}
\institute{Iñaki ECHEVERRÍA-HUARTE \at Departamento de Física y Matemática Aplicada, Facultad de Ciencias, Universidad de Navarra, 31080 Pamplona, Spain \email{iecheverria.13@alumni.unav.es}
\and Alexandre NICOLAS \at Institut Lumière Matière, CNRS \ Université Claude Bernard Lyon 1, 69622 Villeurbanne, France \email{alexandre.nicolas@cnrs.fr}}
%
%

\maketitle

\vspace{-1.5cm}

\abstract{Robust agent-based models for pedestrian dynamics, which can predict the motion of pedestrians in various situations
without specific adjustment of the model or its parameters, are highly desirable. But the modeller's task is challenging, in part because
it mingles different types of processes (cognitive and mechanical ones) and different levels of description (global path planning
and local navigation). We argue that the articulations between these processes or levels are not given sufficient attention in many current modelling frameworks and that this deficiency hampers the effectiveness of these models. Conversely, 
if a decision-making layer and a mechanical one are adequately distinguished, the former controlling the desired velocity
that enters the latter, and if local navigation is not guided solely by intermediate way-points towards the target, but by broader spatial information (e.g., a floor field), then greater robustness can be achieved. This is illustrated with the ANDA model, recently proposed based on such considerations, which was found to reproduce a remarkably wide range of crowd scenarios with a single set of intrinsic parameters.
}

\section{Introduction}
Conventional wisdom has it that the apple doesn't fall far from the tree, and 
 this certainly applies to the modelling of pedestrian dynamics.
Indeed, even as this particularly interdisciplinary field teems with models that come in all
sorts of flavours and with distinct (academic or industrial) goals, one cannot help but notice that these models very generally retain 
distinctive characteristics of their topical roots.

Thus, models originating from the field of algorithmic robotics (or computer graphics) 
are typically focused on the ability to reach a target while avoiding collisions at all costs.
A seminal idea for that purpose has been the introduction of velocity obstacles, i.e., the set of all
velocities leading to a collision before a predefined time horizon \cite{van2008interactive,van2011reciprocal,curtis2013pedestrian}, and the idea that the chosen velocity
should not belong to this set. Model developers from this field are often eager to give mathematical proofs
guaranteeing the absence of collisions, at least in some regimes \cite{karamouzas2017implicit}.

Putting more emphasis on individual choice and less on global maneuverability, economists and econometricians have employed the structure of  discrete-choice models to find which step is optimal \cite{antonini2006discrete} and suitably calibrate their model \cite{robin2009specification}; each agent then chooses to make a step which optimises a utility function depending on various factors. The idea of optimal steps was also taken up in the optimal-step model \cite{seitz2012natural}, from a more pragmatic standpoint.

By contrast, a line of models initially propounded by physicists, first of which the celebrated social force model 
\cite{helbing1995social} and its countless extensions and variants \cite{Chraibi2010generalized,seer2014validating,chen2018social},
keep the formal structure of Newton's second law 
and handle interactions between pedestrians in the same way as mechanical forces. Contacts and collisions between agents are
then possible, particularly at high density, so that these models are frequently used to study evacuations; besides, they 
heavily (arguably, too heavily) rely on these contact forces to reproduce the collective flow of crowds \cite{sticco2020re}.
Moussaid et al. insisted on the more heuristic nature at play in pedestrians' decisions of motion, 
thereby slagging off the deterministic mechanical picture and putting forward simple heuristic rules instead, but they still
resorted to a similar force-based equation in which the heuristic pseudo-force is summed with mechanical forces \cite{moussaid2011simple}.

Rather than siding with either of these modelling branches, we claim that the articulation between cognitive (`heuristic') processes and mechanical forces ought to be considered in greater depth and that much is gained by theoretically revisiting it.
On a different axis, the study of pedestrian mobility requires combining several levels of description, at different length scales. Using the terminology defined in \cite{hoogendoorn2004pedestrian}, even after activities have been scheduled at the
\emph{strategic} level, there remains the need to combine the \emph{operational} dynamics of local navigation with
the route choice operated at the \emph{tactical} level. In practice, a chasm is observed between the last two levels of 
description, which are typically studied by different communities and communicate only via the prescription from the tactical level down to the operational one of a preferred velocity \cite{hoogendoorn2003simulation} or intermediate `way-points' (glocal description) \cite{curtis2013pedestrian}. This articulation is by no means seamless.

\begin{important}{Purpose of this contribution}
In this contribution, we probe these two major articulations and we argue that many difficulties evaporate if the pieces are properly knitted together. With a more partial focus, the ANDA model that we very recently introduced based on these considerations \cite{echeverria2022anticipating} succeeds in replicating surprisingly wide-ranging situations.
\end{important}


\section{Cognitive processes and mechanical layer}

Let us take the physicist's perspective as a starting point. The motion of pedestrian $i$ (as a physical body of mass $m$ and position $\boldsymbol{r}_i $), averaged over a stepping cycle, is governed by Newton's law of motion, viz.,
\begin{equation}
    m \ddot{\boldsymbol{r}}_i = m \frac{\boldsymbol{u}_i^{\star}-\dot{\boldsymbol{r}_i}}{\tau^{\mathrm{mech}}} 
        + \sum_j \boldsymbol{F}_{j \to i}^{\mathrm{c}}
        + \sum_{w \in \mathrm{walls}} \boldsymbol{F}_{w \to i}^{\mathrm{c}},
    \label{eq:Newton_mech}
\end{equation}
where $\boldsymbol{F}_{j \to i}^{\mathrm{c}}$ and $\boldsymbol{F}_{w \to i}^{\mathrm{c}}$ denote contact forces exerted by neighbouring pedestrians and walls, respectively. The first term of Eq.~\ref{eq:Newton_mech}, representing the controllable part of the acceleration \cite{hoogendoorn2003simulation} or
the damped self-propelling force of an active particle (depending on the community that studies it),
indicates that the desired velocity $\boldsymbol{u}_i^{\star}$ is not reached instantly, but only after characteristic time $\tau^{\mathrm{mech}}\approx 0.2\,\mathrm{s}$ in free space, due to the cyclic human gait or the limited friction with the substrate. Importantly,  $\tau^{\mathrm{mech}}$ depends on locomotion and mechanical interactions (being larger, for instance, on a slippery ground or, in a broader context, for a ship compared to ground vehicles), but on no account on the reaction time.

To solve the problem, the mechanical layer centred on Eq.~\ref{eq:Newton_mech} should be coupled to a decision-making layer determining $\boldsymbol{u}_i^{\star}$, which is consistent with the early insight of \cite{hoogendoorn2003simulation}. In conventional force-based models, the desired velocity $\boldsymbol{u}_i^{\star}$ is directed towards an intermediate way-point communicated by the tactical block, with a magnitude depending on the individual, and pseudo-forces are additively inserted into
Eq.~\ref{eq:Newton_mech} to account for the deviations from $\boldsymbol{u}_i^{\star}$ due to the local environment (other agents and walls). Conceptually, this is not satisfactory, because it puts these cognition-mediated effects on the same footing as mechanical forces, in particular subjecting them to the same relaxation time scale $\tau^{\mathrm{mech}}$ for no reason.
This amalgamation is facilitated in practice by the fact that the cognitive reaction time $\tau_{\psi}$ involved in walking is of the same
order of magnitude as $\tau^{\mathrm{mech}}$ and that both (cognitive and mechanical) processes take place within the confines of the same physical entity, the pedestrian. The confusion is much more conspicuous, but of the same nature, if one considers, instead of a pedestrian, a remote-controlled boat, for which $\tau_{\psi} \ll \tau^{\mathrm{mech}}$ and the control operator and the system are spatially separated.

\begin{important}{A different paradigm for the decisional process.}
In reality, the decision-making layer requires a different paradigm than Physics.
Among other options, differential games provide an appealing alternative to handle it \cite{von2007theory,hoogendoorn2003simulation,bonnemain2022pedestrians}. However, since they involve the optimisation of strategies that extend in the future, they are inconvenient to handle. In Section~4, we will show that in many practical situations it may suffice to express the desired velocity as the optimum of an instantaneous utility function (or negative cost), viz.,
\begin{equation}
\displaystyle
    \boldsymbol{u^\star} = \underset{\boldsymbol{u} \in \mathbb{R}^2}{\mathrm{arg min}}\ \EE( \boldsymbol{u}).
\end{equation}
provided that the cost function $\EE$ is adequately chosen to \emph{de facto} include some anticipation over the future.
\end{important}

\section{Articulation between the operational dynamics and the tactical level}
To determine the desired velocity $\boldsymbol{u^\star}$, it is convenient to disentangle local effects from the large-scale aspects entering route choice. To do so, route choice may be computed in a coarse fashion at the tactical level and intermediate goals along this route may then be used as way-points or way-portals \cite{curtis2013pedestrian} for local navigation.

However, we claim that too strict adherence to this partition of tasks leads to several issues.
First, the (tactical) choice of a route is influenced by factors affecting the local navigation, such as congestion due to many 
pedestrians on a path option. Although these factors may be described in a coarse way, for instance via an average density, 
subtle changes that may suddenly favour an option over an alternative, such as the particular motion of a pedestrian on one path,
are thus jettisoned. To clarify this, consider a situation in which an agent faces a dilemma between two almost equivalent paths, one on the left and the other one, on the right, which is less favourable by a very slight margin. Adhering to the above decomposition, the tactical block will select the left path and prescribe a desired velocity oriented to the left. Should a counter-walking 
pedestrian unexpectedly obstruct the left path, the simulated agent will somewhat deviate from their planned local navigation, but still aim for the left path, whereas in reality (s)he would opt for the right path in this circumstance.
In other words, the degeneracy reflecting the quasi-equivalence of two, or more, options is lifted too early because of this decomposition. 
Yet another perspective on this first point is rooted in more abstract considerations: Considering the motion of pedestrians in configuration space (with its prohibited zones due to obstacles) and being given a target region determined at the strategic level, it appears relevant to mathematically define the tactical level as the choice of a homotopy class of trajectories (i.e., a group of trajectories which can be smoothly deformed into each other) and the operational level as the optimisation of the trajectory within this class. Then, avoiding
an (inert or human) obstacle to the right or to the left should belong to the tactical level (because this leads to topologically distinct choices), while this choice is clearly influenced by factors that only pop up at the operational level, such as the exact timing of arrivals. It follows that, even from this abstract perspective, the tactical level and the operational one are necessarily interwoven.
Secondly, coming back to more practical concerns, obstacles of small to moderate sizes are generally handled at the operational level, 
but these interactions, treated similarly to pedestrian collision avoidances, are poorly handled by most current local navigation algorithms when the obstacles have a complex (especially, non-convex) shape.

\begin{important}{Transferring broader spatial information to local navigation.}
To overcome these issues, the idea of the static floor field (which measures the shortest-path distance to the target, possibly taking into account travel discomfort along the path) can be transferred from cellular automata to continuous agent-based models for local navigation. Instead of just storing (at best) the gradient of this field along the optimal path, 
we suggest to store a discretised version of the whole static floor field for each type of agents. In this way, the local
navigation algorithm is fed in with information about the `value' of tentative future positions (by interpolation on the discrete
floor field) and can arbitrate between
different path options in light of the instantaneous local perturbations. For sure, this requires more storage memory, say around a few megabytes per agent type for a $100\,\mathrm{m} \times 100\,\mathrm{m}$ space with a $10\,\mathrm{cm}$ resolution, but
this is no issue at all with any modern computer.
\end{important}

\section{Brief presentation of the ANDA model and numerical simulations}
The previous sections have shed light on conceptual problems in the way in which cognitive processes and mechanical forces, on the one hand, and global path planning and local navigation, on the other hand, are typically handled in pedestrian simulation 
models. Now, we aim to show that models designed to circumvent these conceptual issues are practically very effectual. More
precisely, we will briefly recall the main features of the ANticipated Dynamics Algorithm (ANDA) that we very recently introduced \cite{echeverria2022anticipating}
based on the foregoing considerations and touch on some of its successes in reproducing crowd dynamics; for a detailed presentation of the model and its results, the reader is referred to \cite{echeverria2022anticipating}.

In ANDA, the mechanical layer consists of Newton's second law, Eq.~\ref{eq:Newton_mech}, applied to disks (a crude
approximation of pedestrians' shapes) in two-dimensional space that undergo frictionless Hertzian interactions if they are in contact with one another or with a wall.

\begin{figure}[h]
\begin{center}
\includegraphics[width=5.8cm]{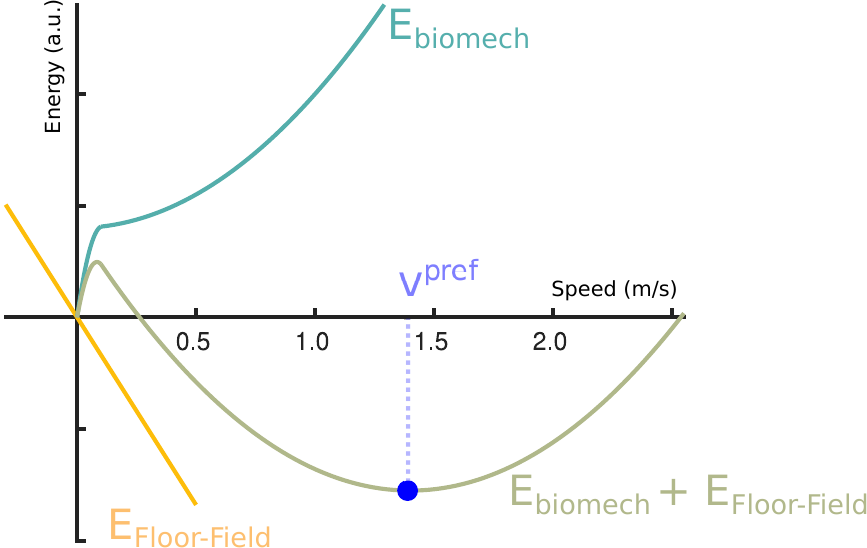}
\sidecaption
\caption{Variations of the bio-mechanical cost $\EE_{\mathrm{biomech}}$ and of the floor field $\EE_{\mathrm{FF}}$ with the test speed $v'$; by definition, the preferential speed $v^{\mathrm{pref}}$ minimises the sum of these two
contributions.}
\end{center}
\label{fig:V_pref}       
\end{figure}

The desired velocity $\boldsymbol{u}_i^{\star}$ entering this equation is obtained from the decision-making layer as the 
velocity that minimises a cost function $\EE[\boldsymbol{u}]$ comprising several contributions,
\begin{center}
\includegraphics[width=0.9\textwidth]{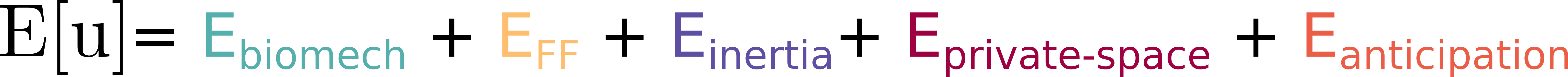}    
\end{center}
In free space, only the bio-mechanical contribution $\EE_{\mathrm{biomech}}$, measuring the (empirical) physiological cost of walking at a
given speed $u= ||\boldsymbol{u}||$, the static floor field $\EE_{\mathrm{FF}}$ exposed in Section~3 (evaluated at
the next position obtained if the test velocity is selected), and the (quadratic)
penalty $\EE_{\mathrm{inertia}}$ of changing velocities too abruptly are operational. In uniform motion, the chosen velocity is then directed along the gradient of $\EE_{\mathrm{FF}}$ and its magnitude $v^{\mathrm{pref}}$ minimises the sum of the first two contributions, as illustrated in Fig.~\ref{fig:V_pref}. Accordingly, if one knows an agent's free-walking speed $v^{\mathrm{pref}}$, the slope of the floor field can directly be obtained and the model contains no
adjustable parameter at this point.

On top of these three contributions, for pedestrians walking \emph{alone} (no groups), interactions with the built environment and the crowd generate two new terms, reflecting
two distinct types of repulsive interactions at play in pedestrian dynamics. The first one, $\EE_{\mathrm{private-space}}$, 
is based on the separation distance between an agent and their neighbours, with a short-ranged repulsive strength decaying with distance, which is familiar to physicists; it reflects the
desire of people to preserve a private space around themselves, whose extent may vary between individuals and between cultures (as studied by the field of proxemics). Beyond these concerns for private space, pedestrians also pay particular attention
to the risk of future collisions and adapt their trajectories to avoid them. Karamouzas et al. demonstrated, using empirical data sets, that these effects are much more readily described using a new variable, the anticipated time to collision (TTC), than distances \cite{Karamouzas2014universal}. More precisely, the TTC is computed as the first time in the future at which a collision is expected if the neighbouring agents keep their current velocities and an expression for the TTC-based interaction
energy was derived by Karamouzas and colleagues, in the form of a cut-off power law. In the ANDA model, we have kept this 
energy $\EE_{\mathrm{anticipation}}$, except that non-physical collisions between private spaces are also taken into account (which results in a smoother profile) and only the most imminent collision is considered.

\begin{figure}[h]
\includegraphics[width=\textwidth]{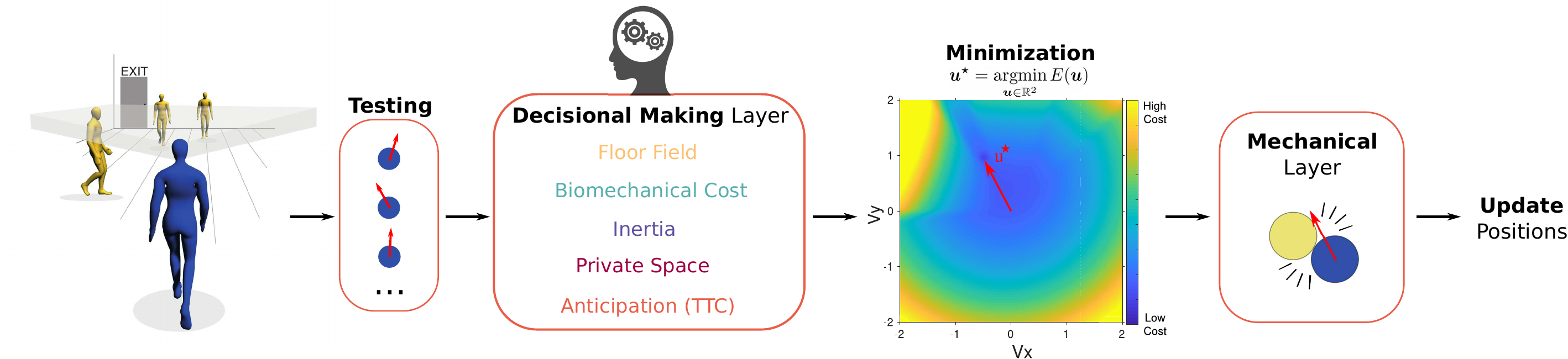}
\caption{General functional diagram of the ANDA model.}
\label{fig:Sketch_model}       
\end{figure}

\begin{figure}[h]
\begin{center}
\includegraphics[width=0.9\textwidth]{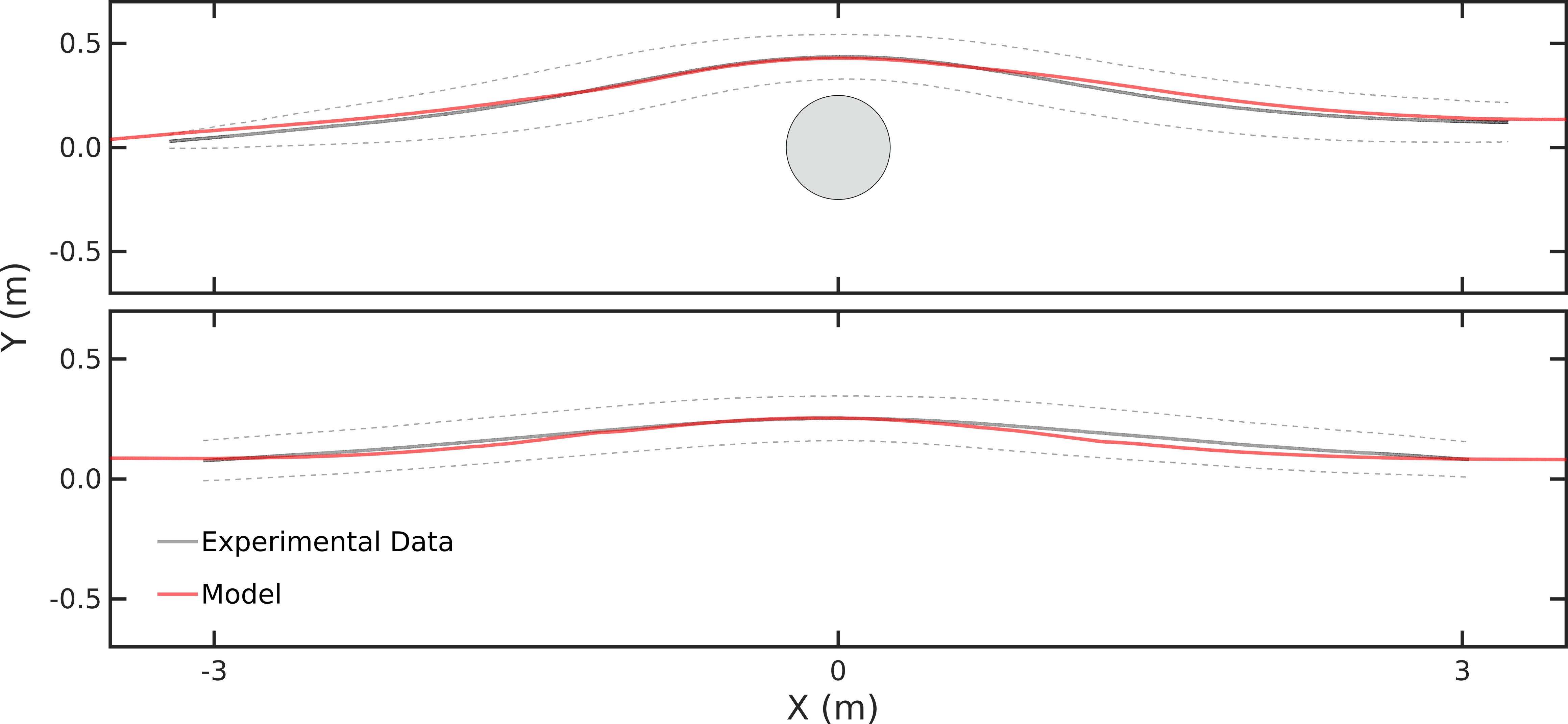}
\end{center}
\caption{Comparison of the mean trajectory along a corridor, averaged over realisations, of a pedestrian avoiding (top) a static pedestrian, (b) a counter-walking pedestrian between controlled experiments of \cite{Moussaid2009experimental} and simulations of the ANDA model.}
\label{fig:Avoidance}       
\end{figure}

Overall, the model follows the functional diagram outlined in Fig.~\ref{fig:Sketch_model} and is implemented in C$^{++}$. There
are only a few parameters (4 to 6, depending on how they are counted) that can be freely adjusted, including  the spatial extent and the strength of the repulsion from the private space and the penality for abrupt velocity changes. Once these are calibrated,
a major success of ANDA is that it can (mostly quantitatively) replicate a very broad spectrum of experimental situations, at various densities; these situations range from binary collision avoidance in a corridor, antipodal motion in which participants initially positioned on a circle have to walk to the diametrically opposed position, unidirectional and bidirectional flow in a corridor, bottleneck flow through a narrow (but not too narrow) doorway, with exit capacities that match experimental data, and
the pedestrian evolution in a complex environment cluttered with obstacles and other pedestrians \cite{echeverria2022anticipating}. As an example, Fig.~\ref{fig:Avoidance} illustrates the results of binary collision avoidance in a corridor. Remarkably, in contrast with previous works, all these situations are reproduced with a unique set of intrinsic model parameters. Ref. \cite{echeverria2022anticipating} also shows that some prominent features associated with digital distraction
by smartphones are correctly captured by the model, if the decisional update time is made longer. 

\begin{svgraybox}
In conclusion, we have argued that the way in which cognitive processes and mechanical interactions are combined in
most pedestrian agent-based models currently in use is theoretically strongly questionable, although restoring the better grounded
sequential articulation wherein a decisional layer feeds a desired velocity into a mechanical equation (as found in \cite{hoogendoorn2003simulation} for instance)
is practically feasible. Besides, the articulation between global path planning and local navigation, generally operated by
defining intermediate way-points, sometimes raises issues that can be alleviated by storing in memory a `tactical' floor field
covering all space. We have exposed that, by adequately handling these articulations in the modelling framework, models can more readily cover a wide range of situations, as exemplified by the ANDA algorithm that was recently introduced
and reproduces quite a broad scope of scenarios with a single set of parameters.
\end{svgraybox}


\end{document}